\documentclass[iop]{emulateapj}

\usepackage{xcolor}
\usepackage{verbatim}
\usepackage{adjustbox}

\shorttitle{Water Emission Toward OH Megamaser Hosts}
\shortauthors{Wiggins, Migenes \& Smidt}

\begin{document}

\title{The Hydroxyl-Water Megamaser Connection. I.  \\
    Water Emission Toward OH Megamaser Hosts}

\author{Brandon K. Wiggins\altaffilmark{1,2} and Victor Migenes\altaffilmark{3}}
\affil{Brigham Young University, Provo, UT 84604}

\and 

\author{Joseph M. Smidt\altaffilmark{4}}
\affil{Los Alamos National Laboratory, Los Alamos, NM, 87544}

\altaffiltext{1}{Graduate Research Assistant, CCS-2, Los Alamos National Laboratory Los Alamos NM}
\altaffiltext{2}{HIDRA Fellow, Brigham Young University}
\altaffiltext{3}{Texas Southern University, Physics Dept., Houston TX}
\altaffiltext{4}{Research Scientist, XTD-IDA, Los Alamos National Laboratory, Los Alamos NM}

\begin{abstract}
Questions surround the connection of luminous extragalactic masers to galactic processes.  The observation that water and hydroxyl megamasers rarely coexist in the same galaxy has given rise to a hypothesis that the two species appear in different phases of nuclear activity. The detection of simultaneous hydroxyl and water megamaser emission toward IC694 has called this hypothesis into question but, because many megamasers have not been surveyed for emission in the other molecule, it remains unclear whether IC694 occupies a narrow phase of galaxy evolution or whether the relationship between megamaser species and galactic processes is more complicated than previously believed.  In this paper, we present results of a systematic search for 22 GHz water maser emission among OH megamaser hosts to identify additional objects hosting both megamaser. Our work roughly doubles the number of galaxies searched for emission in both molecules which host at least one confirmed maser. We confirm with high degree of confidence ($> 8 \sigma$) the detection of water emission toward IIZw96, firmly establishing it as the second object to co-host both water and hydroxyl megamasers after IC694. We find high luminosity, narrow features in the water feature in IIZw96.  All dual megamaser candidates appear in merging galaxy systems suggestive that megamaser coexistance may signal a brief phase along the merger sequence. A statistical analysis of the results of our observations provide possible evidence for an exclusion of H$_2$O kilomasers among OH megamaser hosts.
\end{abstract}

\keywords{masers --- galaxies: evolution ---
galaxies: individual(\objectname{IIZw96},
\object{Arp 299}) --- galaxies: active --- galaxies: starburst}

\section{Introduction}
MASERs (microwave amplification by simulated emission of radiation) are monochromatic, intense radio sources originating from population inversion in clouds of molecular gas. Masers occur astrophysically and are typically associated with star formation, YSOs (e.g. Johanson, Migenes \& Breen 2014; see session 5, IAU Symp. 242, 2007) and late type stars in our local galaxy (e.g. Vlemming et al. 2005). Luminous masers which are observed at extragalactic distances are termed \textit{megamasers}, and widely appear in either hydroxyl (OH; mostly 1665 MHz and 1667 MHz from the hyperfine levels in the ground rotational state $^2 \Pi_{3/2} (J = 3/2)$) or water (22.23508 GHz; $6_{16} \rightarrow 5_{23}$) molecules\footnote{Although the detection of a methanol megamaser has been recently reported, e.g. Chen et al. (2015)}, and are typically associated with mergers, starbursting and active galactic nuclei (AGN) activity (for a review, see Lo 2005). 

Megamasers are remarkable probes of extragalactic phenomena. The parsec-scale environments around supermassive black holes have been probed (e.g. Greenecet al. 2013; Kuo et al. 2011; Reid et al. 2009) and Hubble's constant has been constrained (e.g. Humphreys et al. 2013; Reid et al. 2013; Kuo et al. 2013; Braatz et al. 2010; Herrnstein et al. 1999) through observations of extragalactic water. Hydroxyl (OH) megamasers have been utilized to trace star formation and to estimate the galaxy merger rate (e.g. Darling 2002).  


Unfortunately, megamasers are rare and the precise relationship between megamasers and global galactic processes remains enigmatic. So far, over 4000 galaxies have been searched for 22 GHz water emission with only 150 detections and nearly 500 galaxies have been surveyed for hydroxyl emission with 120 detections (see Wagner 2013 and references therein). Many attempts to link megamasing of either molecular species to other galaxy observables have largely been met with only modest success (e.g. Darling \& Giovanelli 2006; Zhang 2012; Zhu 2011), but significant connections have been achieved, linking OH megamasers to large dense gas fractions (Darling 2007) and water emission to high column densities (Zhang et al. 2006; Castangia et al. 2013), large X-ray luminosities (Kondrakto e al. 2006) and high corrected [OIII]/H$\beta$ (Constantin 2012). Luminous OH emission is frequently powered by starbursting in LIRG/ULIRG galaxies while water megamasers are usually associated with AGN (Lo 2005). 

Though AGN and starbursting activity frequent the same galactic nucleus (e.g. Dixon \& Joseph 2011; Wild et al. 2010), very luminous OH and H$_2$O emission were not until recently observed in the same galaxy (Wagner 2013; Tarchi et al. 2011, hereafter T11).  This observation led to the hypothesis that OH and H$_2$O exclude each other due to the different and respectively restrictive conditions each molecule demands for masing (see Lo 2005, Lonsdale 2002). However, water emission has since been discovered in multiple regions in Arp 299 including Arp 299-A (IC 694), a galactic nucleus also hosting an OH megamaser (T11).  Another possible dual megamaser candidate, IIZw96, was found in a tentative water detection by Wagner (2013). It is thus unclear whether galaxies hosting dual megamasers are experiencing a brief phase of galaxy evolution (such as a transition from starburst nucleus to an AGN) or whether the relationship between the two megamaser species is more complicated than previously believed.

The frequency of simultaneous OH and H$_2$O has previously been somewhat badly constrained as many masing galaxies have yet to be searched for emission in the remaining molecule. Of the nearly 4500 galaxies searched for maser emission in either molecule, $\sim 90$ have been searched for masers of both species (see Tarchi 2012; Wagner 2013). Of those which have been searched and have emission in at least one molecule ($\sim 60$), the majority (38) only emit in water while 18 are only OH emitters.  The number of extragalactic OH megamasers surveyed for H$_2$O emission is thus somewhat under-sampled in the literature. 

Where previous work has attempted to link megamaser luminosities with galaxy observables, ours is the first survey of which we are aware which aims to exclusively address the question of the relationship between the two megamaser species. Here we present results of a survey of known OH megamaser hosts for 22 GHz water emission to probe the dependency of H$_2$O emission on conditions sufficient for OH megamasers.


\section{The Sample}

\begin{figure}
\includegraphics[width = \columnwidth]{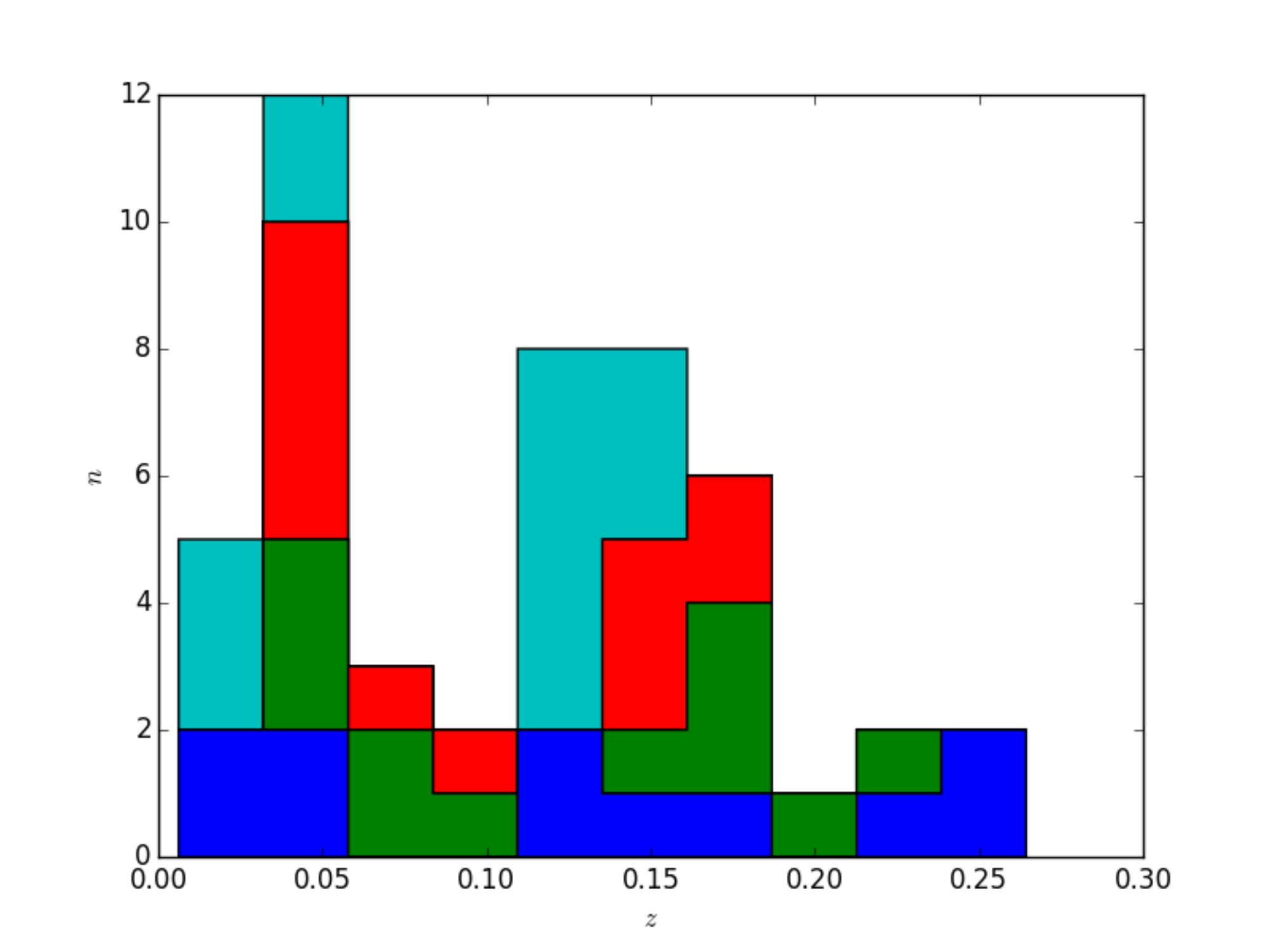}
\caption{The distribution of sample galaxies in redshift $z$ labeled by nuclei type as reported by the NED database. Seyfert-type AGN are in blue, LINER-type AGN appear in green, HII nuclei in red, and galaxies for which no NED classification was available appear in cyan. All galaxies in our survey sample were OH maser hosts. All populations in our sample are relatively well represented in redshift $z$.\\}
\label{ff}
\end{figure}

Our sample is taken from T11 and consists of the subset of their galaxies which are established OH emitters observable by the GBT which had never been observed for H$_2$O emission or for which significantly tighter upper limits on 22 GHz emission might be obtained. Due to observing time limitations, rise times of objects, and RFI, we carried out observations for 22 GHz emission for 47 galaxies hosting OH masers. Figure \ref{ff} summarizes nuclear activity and distribution in redshift $z$ of sample members. Nuclear activity types are fairly well-represented in $z$, mitigating sensitivity biases for detections in a particular group. We note that OH megamasers are fairly uniformly distributed in galaxy activity type (i.e. HII, LINER, Seyfert), occupying roughly the same optical population as (U)LIRG galaxies (Darling \& Giovanelli 2006), so our sample is decently representative of the OH maser/megamaser population.

\section{Observations}

Observations were carried out at the Robert C. Byrd Green Bank Telescope (GBT) between the dates of Feb 27 and April 6, 2014 under project code AGBT14A\_381.  We utilized the GBT spectrometer backend and observed with two 200 MHz bands, the first band being centered on systemic velocity and the second offset by 180 MHz toward lower frequencies. Each band contained 8192 spectral channels giving a velocity resolution of 0.366 km s$^{-1}$ at 20 GHz. Spectra were taken by nodding between beams 3 and 4 of the KFPA receiver every 2.5 minutes for a 5 minute nod cycle.  The telescope was pointed and focused every 30-60 minutes or whenever the telescope slewed more than a total of 10 degrees on the sky. System temperatures were typically 40-60 K. With few exceptions, observations were carried out above elevations of 30$^\circ$ over the horizon with objects closer to the horizon experiencing increases in system temperatures and compromises in RMS sensitivity. 

The project was granted $\sim 60$ hours of telescope time divided into 8 epochs. Good observing conditions and zenith opacities (0.02 - 0.03) were enjoyed over the course of the observations.  All objects were observed in LL and RR circular polarizations which were averaged to produce final spectra. To test our setup, we successfully observed the bright water megamaser in Mrk 1029.

We encountered a bright RFI source from a pair of satellites\footnote{\tiny http://www.spectrumwiki.com/wiki/DisplayEntry.aspx?DisplyId=45, http://www.spectrumwiki.com/wiki/DisplayEntry.aspx?DisplyId=109} which prevented observations of all the galaxies on our source list. We were the first observers to encounter this RFI which likely arose from detecting the side lobes of one or both of these satellites which had recently began servicing nearby locales.  The RFI persisted over our $\sim 1.5$ months of observations. These effects were mitigated by restricting observations to certain portions of the sky.

\section{Data Reduction and Analysis}

Most data reduction was carried out with GBTIDL on NRAO Green Bank computing platforms.  Local weather conditions were determined from the GBT CLEO tool, which estimates zenith opacities by averaging weather data over 3 surrounding stations. Amplitude calibration was carried out with appropriate calibrators in Ott et al. (1998). Spectra were then examined band by band and flagged for obvious signs of RFI.  15 of our 61 observed sources were sufficiently damaged by RFI that no constraints on 22 GHz water emission could be achieved. Each of our nod scans in a given observation (including those across multiple nights of observation) were then averaged. Non-emitting portions of the averaged spectrum were fit with 3rd order polynomial to flatten baselines. During inspection, reference spectra were smoothed with 16 channel boxcar routine to bring out weaker features, though we found that this sometimes introduced spurious spikes in the data. All apparent signals were verified against the spectrum without reference spectral smoothing. All spectra were immediately smoothed with a 4 channel boxcar scheme which was successful in eliminating about half the noise.

Because masers detected in this paper had a distinctly non-gaussian character, line fluxes were first calculated within GBTIDL using their \texttt{stats} procedure, and then subsequently calculated using a numeric integrator based on Simpson's rule. We assume our survey is sufficiently sensitive to detect the flux from a $\sim 3 \sigma$ peak with a FWHM of 2 km s${-1}$ (approximately 5 channels; see Wagner 2013).  Upper limits for isotropic luminosities of masers can then be estimated with the method in Bennert et al. (2009), i.e.

\begin{equation}
\nonumber
\frac{L_{H_2O}}{L_\odot} = \left[0.023 \times \frac{W}{\mbox{Jy} \mbox{ km s}^{-1} }  \right] \times \frac{1}{1+z} \times \left( \frac{D_{L}}{\mbox{Mpc}} \right)^2,
\end{equation}
where $W$ is the sensitivity of the survey which we took as the integrated volume of our $3\sigma$ Guassian curve with a FWHM of 2 km s$^{-1}$. We estimate our errors in calibration to be between $10 - 15$\%.

\section{Results and Discussion}
Our observations are summarized in Table 2. We include all observed sources in this table, including 15 sources badly damaged with RFI, for which upper limits on emission could not be achieved. Note that RMS sensitivities reported are 1 $\sigma$ for the entire band after boxcar smoothing. Of the 46 uncompromised sources observed, we detect confident water emission toward one source and we set new upper limits for water emission on 45 galaxies, 40 of which have yet to be observed for 22 GHz emission. We now discuss our sole maser detection.

\subsection{IIZw96}

\begin{figure}
\includegraphics[width = \columnwidth]{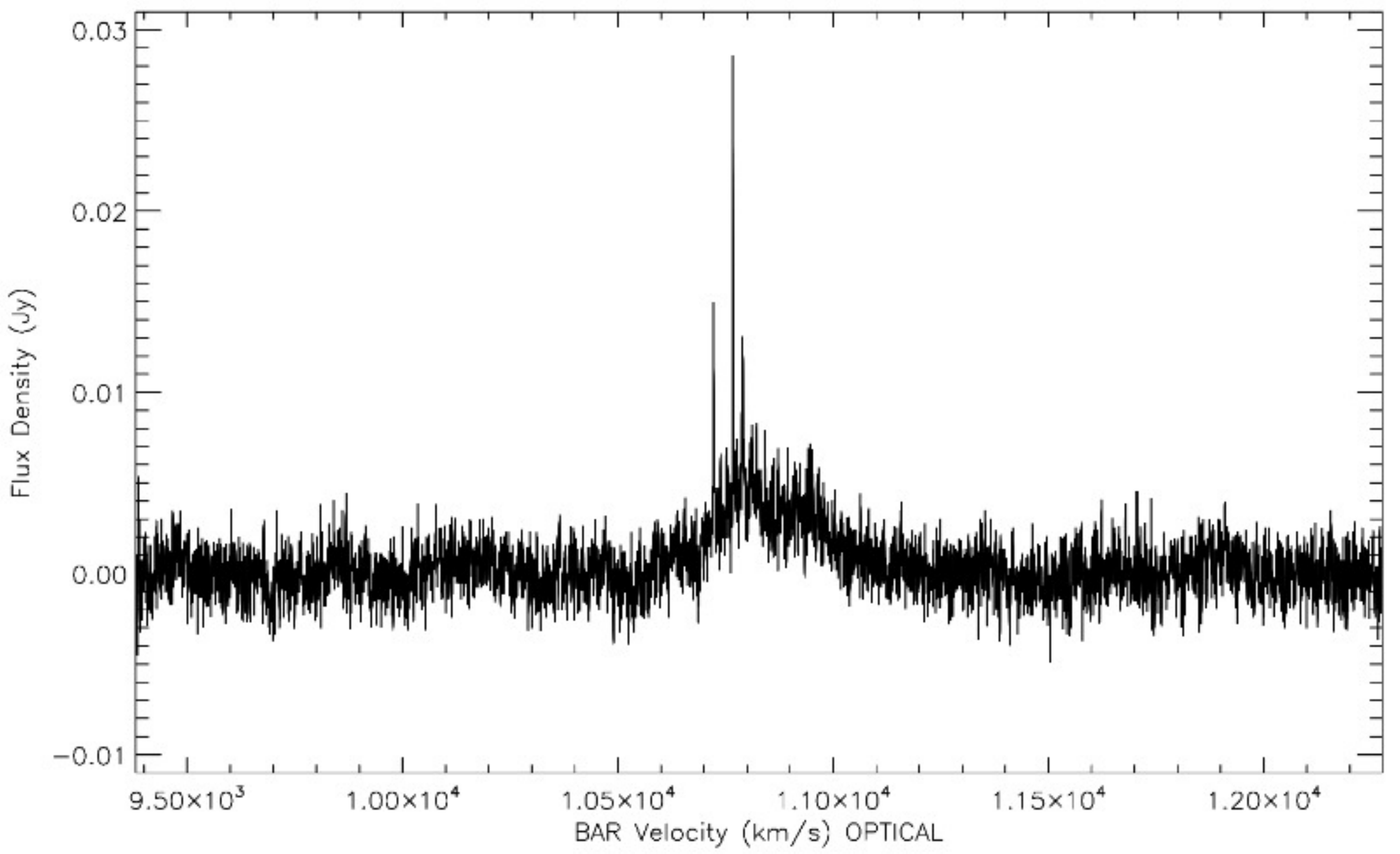}
\caption{22 GHz line detection toward IIZw96 after 1 hour of integration, baseline removal and 4 channel boxcar smoothing. RMS noise is 1.22 mJy.  This figure was generated in GBTIDL.\\}
\label{fig21}
\end{figure}

IIZw96 is a complex merger between at least two galaxies. The OH megamaser is offset from obvious galactic nuclei (Migenes et al. 2012), in a $\sim 10^9 M_\odot$ reddish clump of gas. Soft X-ray emission and strong IR emission lines (Goldader et al. 1997) suggests this clump is an extra-nuclear, obscured starburst, similar to that found in Arp 299, regions C and C' in T11 (Imani et al. 2010). IIZw96 resembles Arp 299 morphologically: Goldader et al. (1997) observes that both galaxies possess similar projected size (10-15 kpc), and are composed of distinct nuclei which are strongly interacting and are in a similar, very short ($\sim 10$ Myr) ``intermediate" stage of merging.

Water emission toward IIZw96 was tentatively detected by Wagner (2013) but could not be confirmed above 5$\sigma$ as integration time was only 12 minutes.  In Figure \ref{fig21}, we present the IIZw96 feature after approximately an hour of integration. This generous observation time was sufficient to resolve the finer features of the water maser which were not previously detected. The $\sim 500$ km s$^{-1}$ profile exhibits a double-humped character in agreement with Wagner (2013) but we do not detect his two side-lobes.  We also obtain an isotropic luminosity somewhat smaller than Wagner (2013) at $\sim 400 L_\odot$ but the maser remains very luminous. This difference might be attributed to Wagner's possible inclusion of tentative side bumps in the maser luminosity in addition to uncertainties in Wagner's measurement arising from short integration times and possibly to short-term maser variability. Our observation confirms this detection beyond doubt ($\sim 8 \sigma$) and establishes IIZw96 as the second system to host formal megamasers in both OH and H$_2$O species.

Ripples in the baseline of the spectrum of IIZw96 persist after baseline removal.  The residual baseline structure is
a time and frequency dependent ripple caused by some combination of RFI and system instability. 

\begin{figure}
\includegraphics[width = \columnwidth]{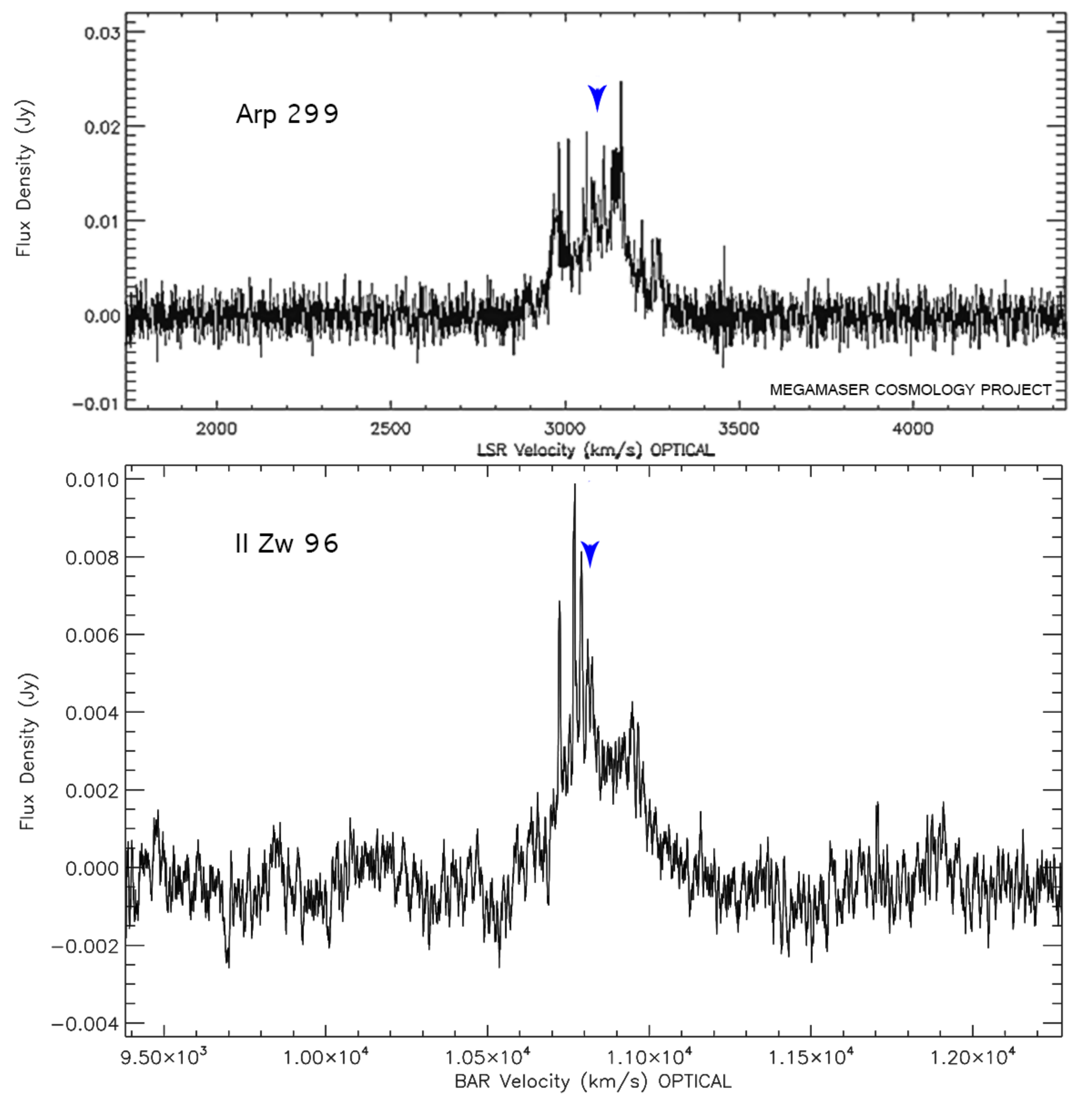}
\caption{\textit{Top Panel:} Water feature in Arp 299 adopted from the Megamaser Cosmology Project (\texttt{https://safe.nrao.edu/wiki/bin/view/Main/ MegamaserCosmologyProject}). \textit{Bottom Panel:} Water feature in IIZw96 after $\sim$ 1 hour integration time and 16-channel boxcar smoothing.  The narrow features on the blue peak reach amplitudes of 28 mJy when 4-channel boxcar smoothing is applied. Systemic velocities are indicated with blue arrowheads.  Note the different scales on vertical axes between panels. The figure was generated in GBTIDL.\\}
\label{fig2}
\end{figure}
              
Wagner (2013) drew connections between IIZw96 and the water feature in NGC 2146 which is associated with star formation. With larger integration times, the line in IIZw96 better resembles the profile of Arp 299 itself (see Figure \ref{fig2}).  We should carefully note, however, that Arp 299's profile contains water emission from several regions within the merger (not just IC 694) and a similar situation could be the case with IIZw96. Because of this and the fact that water emission profiles are time-variable, no conclusive connections between the systems can be made without radio interferometric followups. As with Arp 299, we do not detect high velocity line emission in IIZw96. Our observations reveal narrow features in the water spectrum of IIZw96. Though IIZw96 is among the most powerful starbursts known (Inami et al. 2010), there is a possibility it also hosts an obscured AGN (Migenes et al. 2012) which could provide a pumping mechanism for the maser.

If water emission arises from more than one location in IIZw96, it may make the maser more difficult to detect via interferometric measurements, especially with VLBI, because if the already somewhat low flux is broken up into multiple spatial regions, longer integration times will be required to detect emission from any single region. However, the narrow lines in the water maser emission may suggest the presence of 
compact emission which would render such followup worthwhile.

\section{Effectiveness of the Survey}
A survey for H$_2$O emission among OH megamaser hosts poses a problem in that OH megamasers are spread more widely in $z$, rending maser emission more difficult to constrain. In Figure \ref{fig2}, we plot detection limits of our survey as a function of $z$ for $3 \sigma$ Gaussian peak with 2 km s$^{-1}$ FWHM assuming a 1, 2 and 4 mJy rms noise over the entire 200 MHz band. Superposed on this are the individual upper limits of the masers in our sample, variations in upper limits being derived from system temperatures and differing integration times ($30-60$ minutes). The location of our megamaser detection is indicated with a cross. One will note that at very high $z$ ($z \sim 0.3$), our survey will only be effective in observing masers with $L_{\mbox{\tiny H$_2$O}}/L_\odot \gtrsim 400$.  Though such water masers are inherently rare (see Bennert et al. 2009), very luminous water megamasers at high $z$ have been detected. With only a few exceptions (IRAS 10039-3338, IRAS 11010+4107, IRAS 11506-3851, IRAS 12243-0036, IRAS 15065-1107, IRAS 15247-0945), our survey does not provide upper bounds on water emission below $L/L_\odot = 10$. 

Very luminous water masers tend to be associated with AGN activity and masers appearing on molecular disks may have velocities far (1000s of km s$^{-1}$) displaced from the systemic. At 20 GHz our 200 MHz bands span a velocity of range of $\sim 3000$ km s$^{-1}$ centered on the systemic velocity. Our second 200 MHz band offers similar velocity coverage toward lower frequencies covering a total velocity coverage of about (-4500 km s$^{-1}$, +1500 km s$^{-1}$). In this particular regard, our observations are identical to water observations of e.g. Braatz (2008).  In some cases, RFI prevented full use of the band, so upper bounds on emission were determined from portions of the band near the systemic velocity (see notes on Table 2). In these special cases, our upper bounds do not constrain emission of possible high velocity features. 

Maser luminosities of AGN-pumped water masers can vary dramatically on the timescale of years. We cannot exclude the possibility that a fraction of our nondetections arose from missing maser flares.

Simply integrating the water megamaser luminosity function (Bennert et al. 2009; McKean 2011) over the volume of our survey as characterized by our bandwidth of $\sim 400$ MHz, number of pointings (46) and field of view (2.5 arcmin) with upper and lower isotropic luminosity limits of $10^{-3} L_\odot$ and $10^3 L_\odot$, i.e. assuming the survey is blind, predicts $\sim 3$ masers with luminosities $\geq 1 L_\odot$ in our survey volume.  12\% of 100,000 Monte Carlo simulations which treated maser luminosities with our detection limits predicted we would detect at least one maser. However, with a global detection rate of $\sim  3\%$ across all galaxies surveyed in the literature and $\sim 10\%$ detection rates among Seyfert 2 galaxies, we might estimate the number of detections to be $\sim 1-2$ maser detections with $\sim 1$ detection toward a Seyfert 2 type nucleus.  

\begin{figure}
\includegraphics[width = \columnwidth]{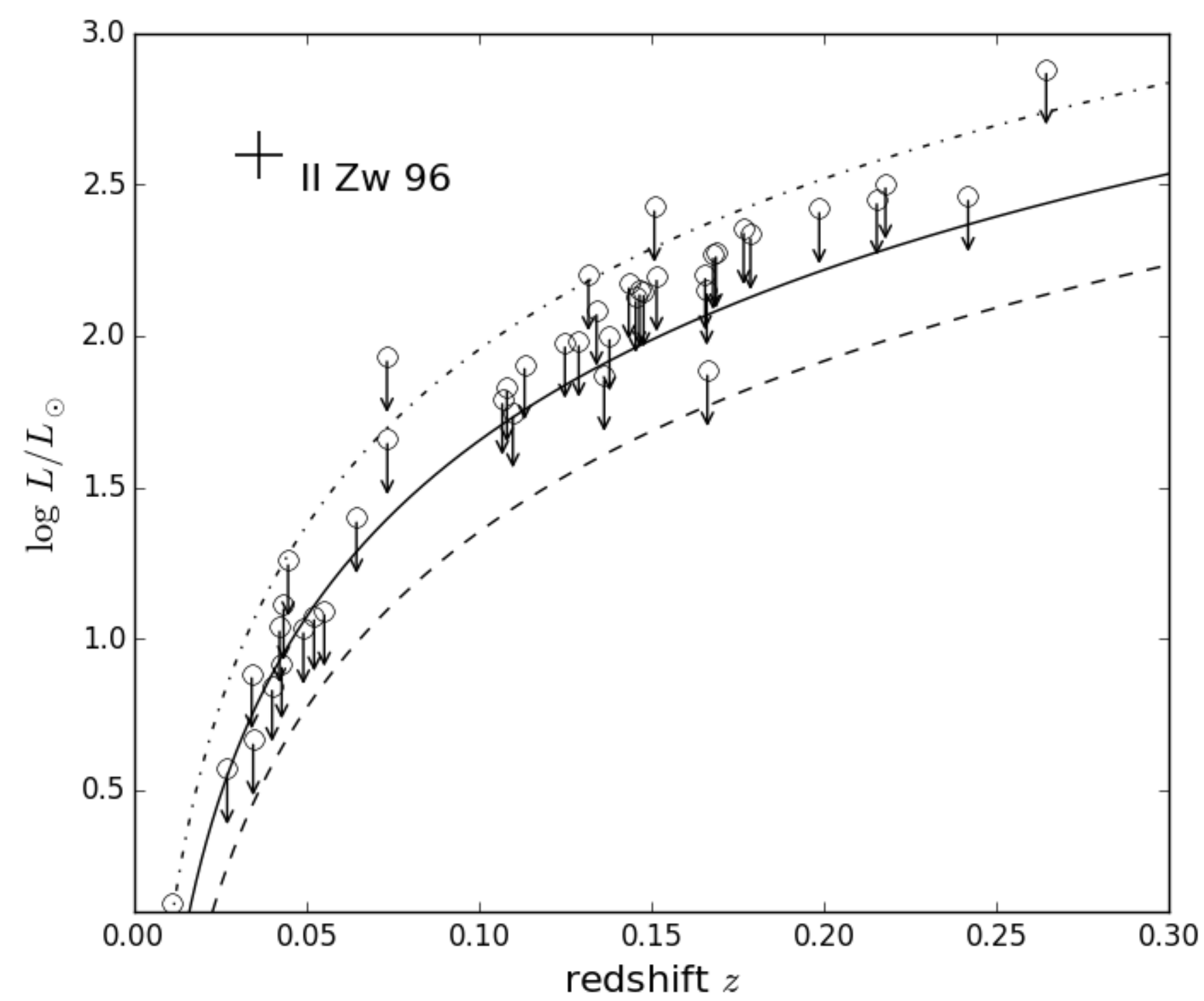}
\caption{Isotropic water maser luminosity upper limits for our survey plotted against redshift $z$. Open circles correspond to H$_2$O non-detections. The dashed, solid and dot-dashed curves are the sensitivity limits of 1, 2 and 4 mJy rms observations for a $\sim 2$ km s$^{-1}$ ($\sim 5.5$ channel) feature. Our survey is only effective at constraining very luminous maser emission at higher $z$. Our maser detection is indicated with a black cross.\\}
\label{fig2}
\end{figure}

\section{The OH and H$_2$O Megamaser Connection}

Previous to this study, approximately 61 galaxies had been searched for both OH and H$_2$O emission which also showed emission in one of the two molecules. With 43 observations of galaxies which have never previously been observed, our study nearly doubles this number. In particular, previous to the present study, 35 objects which had been searched in both molecules showed emission in only water while only 18 galaxies showed only hydroxyl. Now a total of 103 galaxies have been searched for emission in both molecules and show emission in at least one maser species, 60 of these being established OH maser hosts. 

Previous studies have included UGC 5101 as a possible dual megamaser. A very luminous water maser was detected toward UGC 5101 (Martin 1989), which has not been detected in followup observations (Baan et al. 1992). We re-reduced VLBI 1665 MHz observations (VLBA project VLBA\_VSN000352) of UGC 5101 and were unable to either find a line or to image the emission. Of the tentative detections discussed, UGC 5101 is the only maser observation where follow up observations were unsuccessful in detecting a signal. For these reasons, we omit UGC 5101 in our analysis, but will provide some discussion on how our results change if the questionable case of UGC 5101 is considered. 

\begin{figure}
\includegraphics[width = \columnwidth]{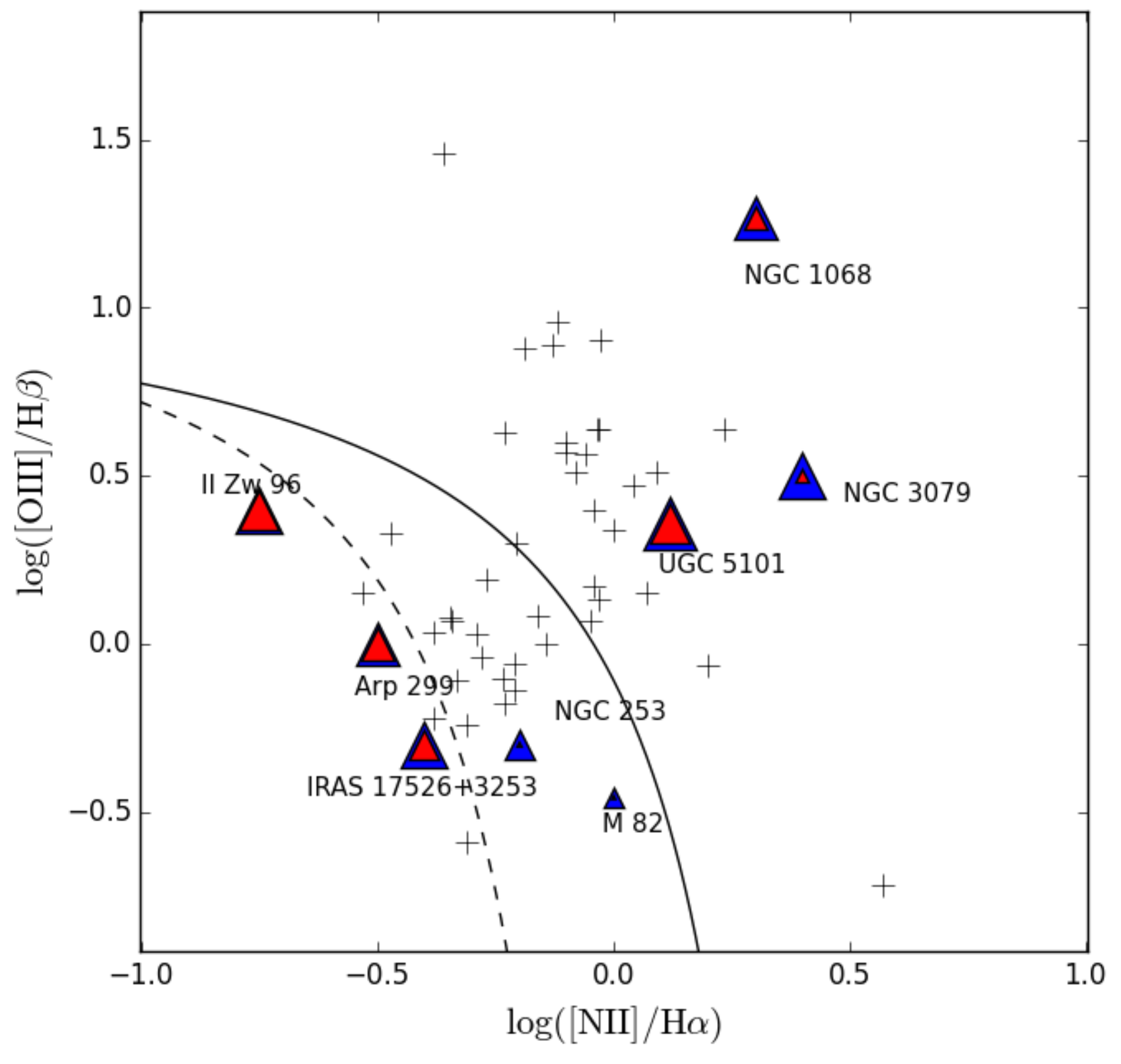}
\caption{
BPT diagnostic tool diagram for as many OH megamaser host galaxies in our survey for which [OIII]/H$\beta$ and [NII]/H$\alpha$ existed in the literature (see text). Plus symbols are OH megamasers for which no water maser emission was detected. Maser emission toward all known dual maser hosts are represented as triangles with the relative size of the triangles proportional to the isotropic luminosities of water (blue) and hydroxyl (red) emission. Note that in all cases of megamaser coexistence, water emission is more luminous than hydroxyl (see discussion in text). Note that fluxes reported are for single dish measurements and that fluxes reported for the entire Arp 299 system should not be confused with fluxes from IC 694 which contains the dual megamaser.\\}
\label{fig3}
\end{figure}

In the bottom panel of Figure \ref{fig3}, we plot a ``Baldwin, Phillips \& Terlevich'' (BPT) diagnostic panel for as many galaxies in our sample as have values for optical spectroscopy line values available in the literature (Lozinskaya et al. 2009; Darling \& Giovanelli 2006; Veillux et al. 1999; Zenner \& Lenten 1993; Garcia-Marlin 2006; Baan et al. 1998). Plus symbols represent OH maser hosts for which no water was detected. In this figure, IRAS 16399-0937 has been included twice as the location of the OH megamaser is unknown (Sales et al. 2014). Blue and inset red triangles mark positions of galaxies with dual masers, the size of the triangles being proportional to $\log[ L/L_\odot]$ for water and hydroxyl respectively. The dashed line is Kauffman et al. (2003) demarcation line distinguishing starburst galaxies from AGN. The solid line is the demarcation criterion from Kewley et al. (2001) or the ``extreme-stardust'' line. With the exception of UGC 5101, both systems (Arp299 and IIZw96) hosting dual megamasers are classified as HII nuclei. In general, however, dual masers trace the BPT positions of sole OH emitters well. A KS test in 
$\log(\mbox{[OIII]})/\mbox{H}\beta$ 
between dual masers and OH maser hosts does not exclude the possibility that the two may be drawn from the same distribution ($p = 0.39$) and a test in 
$\log(\mbox{[NII]})/\mbox{H}\alpha$ 
gives $p = 0.29$. 

\begin{figure*}

\includegraphics[width = \textwidth]{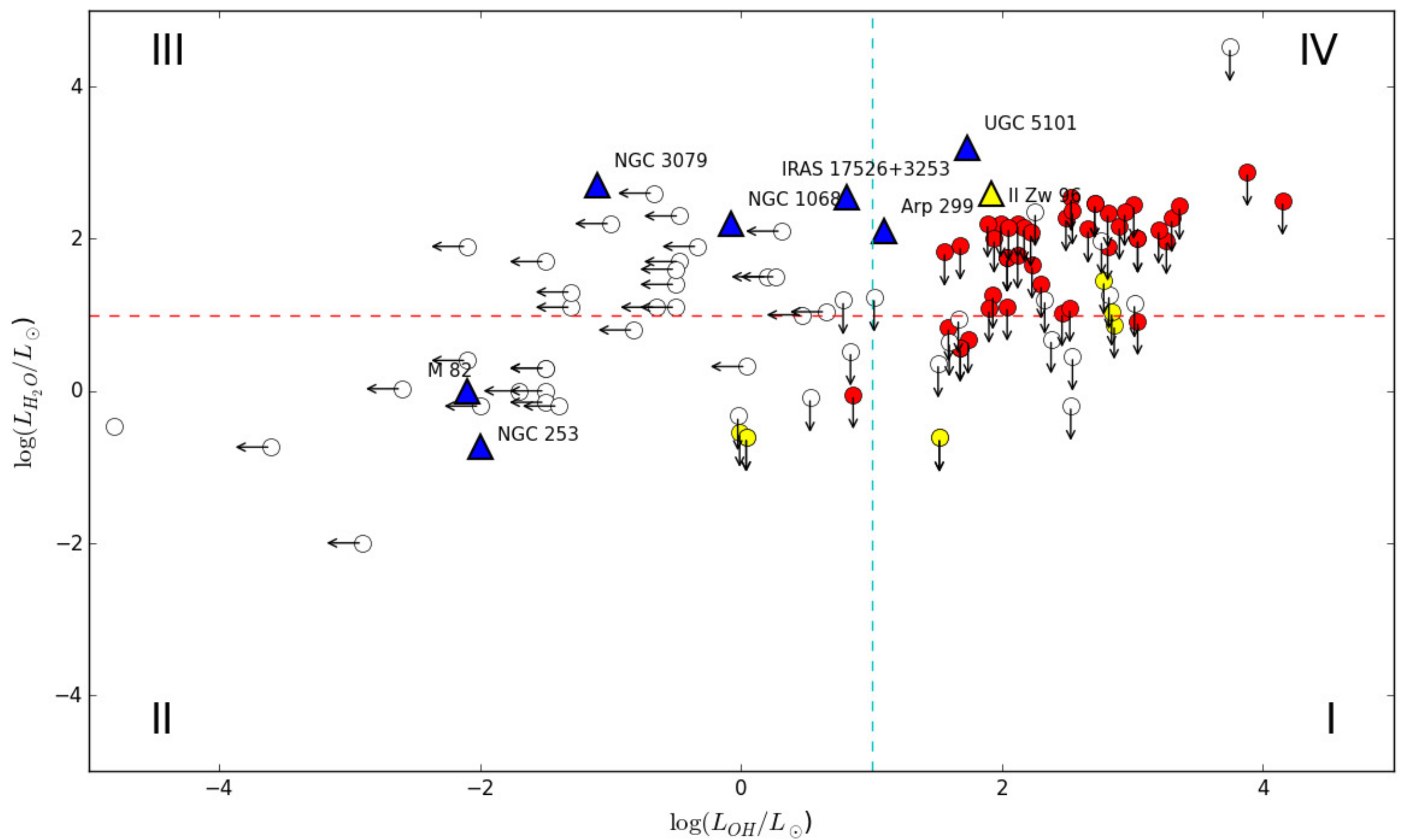}
\caption{OH vs. H$_2$O luminosities for all galaxies searched for both OH and H$_2$O maser emission. As in T11, dual megamaser galaxies appear as triangles. Blue triangles are sources dual megamasers from T11.  Circles or triangles with yellow filling are objects which have been previously observed in the literature (e.g. T11; Wagner 2013) and which were observed again in this study. Symbols with red filling are galaxies which have never been observed previous to this study. The sensitivity of our survey does not eliminate many galaxies as H$_2$O megamaser candidates by Tarchi's criterion ($\log(L_{\mbox{H$_2$O}}/L_\odot) > 1.0$, i.e. many galaxies have upper limits appearing in quadrant IV).\\}
\label{fig4}
\end{figure*}

In Figure \ref{fig4}, we plot all galaxies searched for both OH and H$_2$O emission in OH vs. H$_2$O maser luminosity space. Objects appearing in red are OH megamaser which have never previously been observed in the literature.  Yellow circles corresponding to OH megamasers which have been observed previously but for which tighter upper bounds were achieved in this study. Upper limits in emission in OH and H$_2$O are indicated by horizontal and vertical arrows respectively.

T11 noted a lack of H$_2$O kilomasers in OH megamaser hosts compared to OH kilomasers among H$_2$O hosts, i.e. between Quadrants I (hereafter QI) and QIII. We carried out a survival analysis (e.g. Feigelson \& Babu, 2012; Feigelson \& Nelson 1985) between the heavily censored samples of upper limits and maser luminosities in Q1 and Q3 to investigate a potential of the lack of water kilomasers compared to OH kilomasers in dual megamaser hosts. We used the survival analysis package (Therneau 2015) in R (R Core Team 2013), utilizing both the Mantel-Haenszel and Peto \& Peto tests to bracket the effects of different non-detection weightings. The probabilities that we observe no emission from Q1 with the null hypothesis that there is no difference between maser luminosities between quadrants is $p = 0.0896$ and $p = 0.0863$ for the Mantel-Haenszel and Peto \& Peto tests respectively.  The difference between kilomaser luminosities and hence numbers of detections between the quadrants is thus marginally significant.

In every case of coexistance, OH is less luminous than its H$_2$O counterpart (see Figure \ref{fig3}). Further, OH megamasers coexisting with H$_2$O masers are substantially less luminous than sole OH masers. While this might appear to indicate that coexisting OH megamasers are a distinct population from sole OH megamasers, this is likely due to the independent natures of the water (Bennert et al. 2009; Henkel et al. 2005) and hydroxyl (Darling et al. 2002) luminosity functions.  Dual megamasers are confined to relatively small survey volume compared to the entire OH megamaser catalog.  Comparing luminosity functions reveals that in a given relatively small survey volume\footnote{Here we mean a sufficiently small survey volume such that we do not expect to detect masers with $L > 1000 L_\odot$, where the two luminosity functions intersect. Such luminous OH masers are more abundant per volume than H$_2$O.}, more H$_2$O masers of a given luminosity will appear than OH of the same luminosity which implies that the brightest maser in the volume will likely be of H$_2$O type. Indeed, this is the same effect which makes comparisons between Q1 and QIII of Figure \ref{fig4} difficult. This effect would need to be disentangled before one considers this trend as evidence for an interplay between the two megamaser species.

We find no statistical difference in our water maser detection rate (2.17\%) among OH hosts and the rate of the rest of the literature combined (3.7\%).

So far all galaxies co-hosting dual megamasers, including UGC5101 are merger remnants. If one excludes UGC5101, the remaining two dual megamasers are merging galaxies with spatially distinct nuclei which are optically classified HII-type. Interferometric followups, in particular using VLBI, on IIZw96 is needed to diagnose its masing behavior, but the known additional similarities between IIZw96 and Arp 299 could suggest tight constraints on dual megamasering in general.

\begin{deluxetable}{lllc}
  \tablecolumns{3}
  \tablecaption{Dual Maser Hosts \label{table:masers}}
  \label{tab_properties}
  \tablehead{Name      & $\log(L_{\mbox{\tiny OH}}/L_\odot)$ & $\log(L_{\mbox{\tiny H$_2$O}}/L_\odot)$ & Type}
  \startdata
   Arp 299 & 1.1 & 2.1 & HII \\
   IIZw96 & 1.9 & 2.6 & HII \\
   NGC 253 & -2.0 & 0.74 & HII / Sy2\\
   NGC 1068 & -0.08 & 2.2 & Sy2 \\
   \textit{UGC 5101} & 1.73 & 3.2 & Sy1 \\
   M 82 & -2.1 & 0.0 & HII \\
   NGC 3079 & -1.1 & 2.7 & Sy2 \\
   \textit{IRAS 17526} & 0.81 & 2.55 & HII \\
  \enddata
  \tablecomments{Maser luminosities adopted from this work, T11 and Wagner (2013). Italicized entries have been reported in the literature as tentative detections. IIZw96 was confirmed in this study. Galaxy classifications were obtained from the NED database.}
    \end{deluxetable}

\section{Conclusions}
We have reported on a systematic search for 22 GHz water emission in established OH megamaser hosts with the Robert C. Byrd Green Bank Telescope (GBT). Our $\sim 60$ hour survey of 47 sources nearly doubles the number of galaxies now searched for both OH and H$_2$O emission with confirmed emission in at least one molecule. This effort is additionally significant because many galaxies searched for emission in both molecules are established water masers leaving many OH megamasers un-probed.  We confirm to $>8\sigma$ the previously tentative water maser detection toward IIZw96.  This finding reinforces the celebrated similarities (e.g. Imani et al. 2010; Goldader et al. 1997) between IIZw96 and Arp 299, the only other known host of both megamaser species.  Our hour-long integration was sufficient to resolve multiple, luminous narrow water features toward IIZw96.  IIZw96 provides another vital site to investigate dual megamasing.

Our survey is sensitive enough to eliminate 6 galaxies as H$_2$O megamaser ($L/L_\odot > 10$) candidates. For the first time, we verify, on a statistical basis, a marginally significant lack of H$_2$O kilomasers among OH megamaser hosts. The two, and so far only, dual megamaser hosts are HII type and are galaxy mergers with spatially distinct nuclei.

\acknowledgments

B.W. was supported by the High Impact Doctoral Research Award (HIDRA) at Brigham Young University. We would like to thank Dr. Jim Braatz and Dr. Amanda Kepley at GBT and Dr. Jennifer Donley at LANL for their counsel and useful conversations during the course of this project.  We would also like to thank our thoughtful anonymous referee whose insightful comments which have greatly improved the quality of this work.

Facilities: \facility{GBT}, \facility{LANL}

\clearpage

\clearpage

\clearpage

\tiny

\begin{deluxetable}{lcrcccrl}

\tabletypesize{\small}

\tablecaption{Megamaser Sample and Summary of Observations}\label{tab:a}
\tablewidth{0pt}
\tablehead{
\colhead{object name} & \colhead{RA} & \colhead{DEC} & \colhead{$z$} & 
\colhead{rms$^*$ \tiny (mJy)} & \colhead{\tiny $\log(L_{OH}/L_\odot)$} &
\colhead{\tiny $\log(L_{H_2O}/L_\odot)$} & \colhead{Notes}
}
\startdata
IRAS 00057+4021 & 00:08:20.5 & +40:37:57 & 0.044660 &  3.77 &  1.93 & $< $1.26 & Sy 2  \\
IRAS 03260-1422 & 03:28:24.3 & -14:12:07 & 0.043350 & 2.78 &  2.04 & $< $ 1.11 & Starburst\\
IRAS 03521+0028$\ddagger$ & 03:54:42.2 & +00:37:03 & 0.151910 &   -  & 1.96 & - & Starburst\\
IRAS 03566+1647$\ddagger$ & 03:59:29.1 & +16:56:26 & 0.133522  &- & 1.61 & -& Sy2\\
IRAS 04121+0223$\ddagger$  & 04:12:47.1 & +02:30:36 & 0.122424 &  -  & 2.29 & -& HII \\
IRAS 06487+2208 & 06:51:45.8 & +22:04:27 & 0.143390 &  2.70 & 2.90 & $< $2.17 & HII, ULIRG\\
IRAS 07163+0817$\ddagger$ & 07:19:05.5 & +08:12:07 & 0.110973 &- & 1.47 &  - & HII\\
IRAS 07572+0533$\ddagger$ & 07:59:57.2 & +05:25:00 & 0.190000 &- & 2.63 & -& LINER, ULIRG\\
IRAS 08071+0509 & 08:09:47.2 & +05:01:09 & 0.052203 & 1.66 &  1.90 & $< $ 1.08 & Radio Jet, HII \\
IRAS 08201+2801 & 08:23:12.6 & +27:51:40 & 0.167830 &  2.44 &  3.30  & $< $2.27 & HII, ULIRG\\
IRAS 08279+0956$\ddagger$ & 08:30:40.9 & +09:46:28 & 0.208634 &- & 2.98 & -& LINER \\
IRAS 08449+2332 & 08:47:50.2 & +23:21:10 & 0.151458 &  2.54 &  2.12 & $< $2.20 & HII, ULIRG\\
IRAS 08474+1813$\dagger$ & 08:50:18.3 & +18:02:01 & 0.145404 &  2.35 &    2.66 & $< $ 2.13 & Sy 2, ULIRG\\
IRAS 09039+0503$\ddagger$ & 09:06:34.2 & +04:51:25 & 0.125140 & -& 2.63 & -& LINER\\
IRAS 09531+1430$\dagger$ & 09:55:51.1 & +14:16:01 & 0.215275 & 2.20 &  3.01 & $< $2.45 & Sy 2\\
IRAS 09539+0857$\dagger$ & 09:56:34.3 & +08:43:06 & 0.128899 &  2.14 &  3.26  & $< $1.98 & \\
\textit{IRAS 10039-3338} & 10:06:05.1 & -33:53:17 & 0.034100 &  2.47 &  2.86 &  $< $ 0.87  & LIRG, merger \\        
IRAS 10036+2740 & 10:06:26.3 & +27:25:46 & 0.165531 &  2.14 & 1.99   &  $< $ 2.204 & ULIRG \\
IRAS 10173+0828 & 10:20:00.2 & +08:13:34 & 0.049087 &  1.67 &  2.46 &  $< $ 1.03 &LIRG \\
IRAS 10339+1548$\ddagger$ & 10:36:37.9 & +15:32:42 & 0.197236 & -& 2.34 & -& Sy 2, ULRIG\\
IRAS 10378+1109$\dagger$ & 10:40:29.2 & 10:53:18 & 0.136274 &  2.61 &  3.20 & $< $2.12 & ULIRG, LINER\\
IRAS 11010+4107 & 11:03:53.2 & +40:50:57 & 0.034524 &  1.47 &  1.74 & $< $0.67 & LIRG, ring-like\\
IRAS 11029+3130 & 11:05:37.5 & +31:14:32 & 0.198604 &  2.43 &  2.53 & $< $2.55 & LINER, ULIRG\\
IRAS 11180+1623 & 11:20:41.7 & +16:06:57 & 0.166000 &  1.89 &  2.17 & $< $2.15 & LINER, ULIRG\\
\textit{IRAS 11506-3851} & 11:53:11.7 & -39:07:49 & 0.010781 &      3.84 & 1.52 & $< $-0.613 & LIRG\\
IRAS 11524+1058 & 11:55:02.8 & +10:41:44 & 0.178670 &  2.48 & 2.81 & $< $ 2.34 & LINER, ULIRG\\
IRAS 12018+1941 & 12:04:24.5 & +19:25:10 & 0.168646 &  2.43 &  2.49 & $< $ 2.28 & Sy 2\\
IRAS 12032+1707 & 12:05:47.7 & +16:51:08 & 0.217787 &  2.42 &  4.15 & $< $2.50 & LINER, ULIRG\\
\textit{IRAS 12112+0305} & 12:13:46.0 & +02:48:38 & 0.073317 &  5.94 & 2.78 & $< $1.454  & LINER, ULIRG, HII \\ 
IRAS 12162+1047 & 12:18:47.7 & +10:31:11 & 0.146500 &  2.43 &  2.05 & $< $ 2.15 & \\
\textit{IRAS 12243-0036} & 12:26:54.6 & -00:52:39 & 0.007268 &  1.60 &  -0.01 & $< $ -0.546 & Sy 2, LIRG\\ 
\textit{IRAS 12540+5708} & 12:56:14.2 & +56:52:25 & 0.042170 &  2.37 &  2.84 & $< $1.04 & LIRG, Sy 1\\ 
IRAS 12548+2403 & 12:57:20.0 & +23:47:46 & 0.131700 & 3.40 &  1.89 & $< $ 2.20 & \\
IRAS 13218+0552$\ddagger$ & 13:24:19.9 & +05:37:05 & 0.205102 &- & 3.07 & -& Sy 1.5\\
IRAS 14043+0624$\ddagger$ & 14:06:49.8 & +06:10:36 & 0.113187 &- & 1.68 &- &  \\
IRAS 14059+2000$\ddagger$ & 14:08:18.7 & +19:46:23 & 0.123700 &- & 2.97 &- & \\
IRAS 14070+0525 & 14:09:31.2 & +05:11:31 & 0.264380 &  3.88 &    3.88 &$< $2.88 & ULIRG, Sy 2\\
IRAS 14553+1245 & 14:57:43.4 & +12:33:16 & 0.124900 &  2.28 &  1.94 & $< $ 2.00 & \\
IRAS 14586+1431 & 15:01:00.4 & +14:20:15 & 0.147700 &  2.39 & 2.54 & $< $2.37 & \\
\textit{IRAS 15065-1107} & 15:09:16.1 & -11:19:18 & 0.006174 &  2.10 &  0.04 & $< $-0.61 & Sy 2, HII\\
\textit{IRAS 15179+3956}$\ddagger$ & 15:19:47.1 & 39:45:38 & 0.047570 & - & 1.02 & $<$1.23$^{**}$ &  merger, HII \\ 
IRAS 15224+1033 & 15:24:51.5 & +10:22:45 & 0.134049 &  2.53 &    2.22 & $< $2.08 & \\
IRAS 15250+3609 & 15:26:59.4 & +35:58:38 & 0.055155 &  1.57 &  2.52 & $< $1.09 & ULIRG, LINER\\
IRAS 15247-0945 & 15:27:27.8 & -09:55:41 & 0.040004 &  1.67 & 1.59 & $< $0.84 & LINER\\
IRAS 15587+1609$\dagger$ & 16:01:03.6 & +16:01:03 & 0.137181 &  2.00 & 3.04 & $< $2.00 & HII\\
IRAS 16100+2528$\ddagger$ & 16:12:05.4 & +25:20:23 & 0.132385 & - & 1.68 & -& LINER\\
IRAS 16255+2801$\ddagger$ & 16:27:38.1 & +27:54:52 & 0.133612 & - & 2.38 & -& HII, ULRIG\\
IRAS 16300+1558 & 16:32:21.4 & +15:51:46 & 0.241747 &  1.79 & 2.71 & $< $ 2.46 & ULIRG, Sy2\\
IRAS 16399-0937 & 16:42:40.2 & -09:43:14 & 0.027012 &  2.00 &  1.68 & $< $0.570 & LIRG\\ 
IRAS 17160+2006 & 17:18:15.6 & +20:02:58 & 0.109800 & 1.77 &    2.04 & $< $1.75 & \\
IRAS 17208-0014 & 17:23:21.9 & -00:17:01 & 0.042810 &  1.80 &     3.04 & $< $0.92 & Starburst, LINER\\
IRAS 17540+2935 & 17:55:56.1 & +29:35:26 & 0.108108 &  2.23 &  1.56 & $< $1.83 & LINER\\
IRAS 18368+3549 & 18:38:35.4 & +35:52:20 & 0.116170 &  2.28 &   2.81 &$< $1.90 & Sy 2\\
IRAS 18544-3718 & 18:57:52.7 & -37:14:38 & 0.073424 &  3.32 &  2.23 & $< $1.66 & HII, Spiral\\
IRAS 18588+3517 & 19:00:41.2 & +35:21:27 & 0.106727 &  2.09 & 2.12 & $< $1.79 & HII, Spiral\\
\textit{\textbf{IRAS 20550+1656}} & 20:57:23.9 & +17:07:39 & 0.036098 &  1.28 & 1.92 & \textbf{2.60} & HII, LIRG, merger\\
IRAS 21077+3358 & 21:09:49.0 & +34:10:20 & 0.176699 &  2.71 &  2.94 & $< $ 2.35 & LINER\\
IRAS 21271+2514 & 21:29:29.4 & +25:27:50 & 0.150797 &  2.80 &  3.36 & $< $2.43 & \\
IRAS 22025+4205 & 22:04:36.1 & +42:19:38 & 0.014310 &  2.02 & 0.86 & $< $ -0.047 & Spiral\\
IRAS 23019+3405$\ddagger$ & 23:04:21.1 & +34:21:48 & 0.108038 & - & 1.78 & - & Sy 2\\
IRAS 23365+3604 & 23:39:01.3 & +36:21:08 & 0.064480 &  2.44 & 2.30 & $< $1.40 & pec, LINER, ULIRG\\
\enddata

\tablecomments{ Boldfaced entries represent detections. Italicized objects have been observed previously, but this study provies tighter upper limits for water megamasers.}

\tablenotetext{}{\footnotesize Galaxy classifications, coordinates and redshifts were adopted from NED database.}
\tablenotetext{$*$}{\footnotesize 1 $\sigma$ RMS noise for the entire observing band reported after 4 channel boxcar smoothing.  Channel widths are $\sim 0.5$ km s$^{-1}$.}
\tablenotetext{$**$}{\footnotesize Upper bound taken from Wagner (2013).}
\tablenotetext{}{\footnotesize Hydroxyl megamaser luminosities adopted from T11 and Wagner (2013).}
\tablenotetext{$\dagger$}{\footnotesize RFI damaged portion of spectrum at frequencies displaced more than 500 km/s from systemic velocity. This constraint assumes no high-velocity features.}
\tablenotetext{$\ddagger$}{\footnotesize Source for which no constraints on H$_2$O emission could be achieved due to RFI.}

\end{deluxetable}
\end{document}